\documentstyle[prc,aps,epsf,twocolumn,eqsecnum]{revtex}

\catcode`\@=11

\def\maketitle2{\par 
\begingroup
\let\cite\@bylinecite
\def\thefootnote{\fnsymbol{footnote}}%
\twocolumn[\@maketitle2\vskip2pc]%
\thispagestyle{plain}\@thanks
\endgroup
\def\thefootnote{\arabic{footnote}}%
\setcounter{footnote}{0}%
\let\maketitle2\relax \let\@maketitle2\relax
\let\@thanks\relax \let\@authoraddress\relax \let\@title\relax
\let\@date\relax \let\thanks\relax \let\@abstract\relax 
\let\@pacs\relax}

\def\abstract#1{\gdef\@abstract{{\par 
\bgroup
\ifdim\prevdepth=-1000pt \prevdepth0pt\fi
\hsize\columnwidth
\dimen0=-\prevdepth \advance\dimen0 by17.5pt \nointerlineskip
\small\vrule width 0pt height\dimen0 \relax}{~~}#1\egroup}}

\def\pacs#1{\gdef\@pacs{{\par 
\bgroup
\hsize\columnwidth \parindent0pt
\ifdim\prevdepth=-1000pt \prevdepth0pt\fi
\dimen0=-\prevdepth \advance\dimen0 by20pt\nointerlineskip
\egroup} PACS numbers:~#1}}

\def\@maketitle2{
\@preprint
\@title
\ifdim\prevdepth=-1000pt \prevdepth0pt\fi
\@authoraddress
\@date
\begin{list}{}{\leftmargin=0.10753\textwidth \rightmargin=\leftmargin
\itemsep=1pc\partopsep=-1pc}
\item\@abstract
\item\@pacs
\end{list}
}

\catcode`\@=12

\begin{document}
\draft

\title{A Measurement of the Interference Structure Function, 
$R_{LT}$, for the $^{12}C(e,e'p)$ reaction in the Quasielastic 
Region.}

\author{
M.~Holtrop$^{1,}$\cite{UNH},
D.~Jordan$^{1,}$\cite{TRIUMF},
T.~McIlvain$^{1}$,
R.~Alarcon$^{2}$,
R.~Beck$^{3}$,
W.~Bertozzi$^{1}$,
V.~Bhushan$^{1}$,
W.~Boeglin$^{1,}$\cite{FIU},
J.P.~Chen$^{1,}$\cite{CEBAF},
D.~Dale$^{1,}$\cite{UKL},
G.~Dodson$^{1}$,
S.~Dolfini$^{3}$,
K.~Dow$^{1}$,
J.~Dzengeleski$^{1}$,
M.B.~Epstein$^{4}$,
M.~Farkhondeh$^{1}$,
S.~Gilad$^{1}$,
J.~G\"{o}rgen$^{2}$,
K.~Joo$^{1,}$\cite{UVA},
J.~Kelsey$^{1}$,
W.~Kim$^{3,}$\cite{KNU},
R.~Laszewski$^{3}$,
R.~Lourie$^{5,}$\cite{SUNY},
J.~Mandeville$^{3,}$\cite{MGH},
D.~Margaziotis$^{4}$,
D.~Martinez$^{2}$,
R.~Miskimen$^{6}$,
C.~Papanicolas$^{3,}$\cite{UATH},
S.~Penn$^{1,}$\cite{UW}
W.~Sapp$^{1}$,
A.J.~Sarty$^{7}$,
D.~Tieger$^{1}$,
C.~Tschalaer$^{1}$,
W.~Turchinetz$^{1}$,
G.~Warren$^{1}$,
L.~Weinstein$^{1,}$\cite{ODU},
S.~Williamson$^{3}$
}
\address{
$^{1}$Massachusetts Institute of Technology, Cambridge, Massachusetts 02139\\
$^{2}$Arizona State University, Tempe, Arizona 85287\\
$^{3}$Nuclear Physics Laboratory, University of Illinois at Urbana-Champaign, 
Champaign, Illinois 61820\\
$^{4}$California State University at Los Angles, 
Los Angeles, California 90032\\
$^{5}$University of Virginia, Charlottesville, Virginia 22901\\
$^{6}$University of Massachusetts at Amherst, Amherst, Massachusetts 01003\\
$^{7}$Florida State University, Tallahassee, Florida 32306\\
}
\date{\today}

%
%
\abstract{
The coincidence cross-section and the interference structure function,
$R_{LT}$, were measured for the $ ^{12}C(e,e'p)^{11}B $ reaction at
quasielastic kinematics and central momentum transfer of $ |\vec{q}\,|
=400\;{MeV}\!/{c}$ . The measurement was at an opening angle of
$\theta_{pq}=11^{\circ}$, covering a range in missing energy of $E_{m} = 0$
to $65 MeV$. The $R_{LT}$ structure function is found to be consistent with
zero for $E_{m} > 50 MeV$, confirming an earlier study which indicated that
$R_{L}$ vanishes in this region. The integrated strengths of the p- and
s-shell are compared with a Distorted Wave Impulse Approximation
calculation. The s-shell strength and shape are compared with
a Hartree Fock-Random Phase Approximation
calculation. 
The DWIA calculation overestimates the cross sections for
p- and s-shell proton knockout as expected, but surprisingly agrees 
with the extracted $R_{LT}$ value for both shells. 
The HF-RPA calculation describes the data more
consistently, which may be due to the inclusion of 2-body currents in
this calculation.  
}

%
%

\pacs{25.30.Dh, 21.10Jx, 21.60Jz}

\maketitle2
\narrowtext

\section{Introduction}

The coincidence $(e,e'p)$ reaction in the quasielastic region
$\left( {\omega \approx {{Q^2} \mathord{\left/ {\vphantom {{Q^2}
{2M_p}}} \right. \kern-\nulldelimiterspace} {2M_p}}} \right)$ 
has long been recognized as an ideal tool to study the single particle
wave-functions of the proton in the nucleus, and such experiments have
resulted in quantitative tests of the nuclear mean field theory\cite{1}. 
Until recently, most experiments only measured cross-sections.
This experiment also extracts one of the structure functions.  The
separation of structure functions provides additional information
about the electron scattering process.  These structure functions are
sensitive to specific aspects of the reaction mechanism or to
properties of the target nucleus (final state interactions, meson
exchange currents, multi-body effect, wave function, nucleon form
factors...), and their measurement will be crucial for the
understanding of these detailed aspects of the electron scattering
process\cite{2}\cite{Lapikas}.

The quasielastic cross-section was originally attributed entirely to
single particle knockout\cite{3}, but more recent experiments suggest
that a considerable amount of strength comes from two-body and
possible multi-body processes\cite{4,5}.  Evidence for multinucleon
processes is also seen in $(e,e'p)$ experiments on carbon in the dip
and delta kinematic regions \cite{6,7,Kester,Kester2}. 
These studies show that there
is more strength at large missing energies than is expected from a
one-body process, and that the transverse part of the cross-section is
enhanced compared to the longitudinal part. A similar result is seen
in an L/T separation for the $^{3}He(e,e'p)$ reaction\cite{8}. An 
experiment on $^{16}O$ for p-shell knockout showed an enhancement of the
$R_{LT}$ structure function\cite{Spaltro}. These
anomalies have not been explained by either final state interactions
or meson exchange currents. A number of previous $R_{LT}$ measurements
were performed on deuterium \cite{Jordan,Ducret,Kasdorp,Frommberger} where
multi-body processes do not play a role.
The measurement reported here is aimed at
providing further information regarding the character of this excess
strength through the isolation of the $R_{LT}$ response function. The
kinematics for the extraction of $R_{LT}$ were chosen to be similar to
ref \cite{4}, where a signature of the anomalous strength has already
been observed in a separation of $R_{L}$ and $R_{T}$.

In the one-photon exchange approximation, the interaction of the
electron with the target nucleus can be described by the exchange of a
single virtual photon having a four-momentum
$q_\mu =\left( {\omega ,\vec{q}\,} \right)$. The spin 
averaged cross-section for the $(e,e'p)$ reaction 
can then be expressed as\cite{9}:
\begin{eqnarray}
{{d^5\sigma } \over {dE_{f}d\Omega _{e}d\Omega _{p}}}&=&{1 \over {\left(
  {2\pi } \right)^3}}C_{kin}\sigma _{M} f_{rec}^{-1}\times   \hfill\cr
  & & \left\{ \matrix{ v_{L}R_{L}+v_{R}R_{R}+ \hfill\cr
   v_{TT}R_{TT}\cos \left( {2\phi _{pq}}\right)+ \hfill\cr\
   v_{LT}R_{LT}\cos \left( {\phi _{pq}} \right)  \hfill\cr } \right\}
\label{eq:cross}
\end{eqnarray}
where the $v_{\alpha\beta}$ terms represent the lepton tensor elements
which depend only on the electron vertex, and the $R_{\alpha\beta}$
terms represent the nuclear response functions. The indices denote
longitudinal ($L$) or transverse ($T$) polarization of the virtual
photon with respect to the momentum transfer direction. Furthermore,
$\phi_{pq}$ is the angle between the electron scattering plane
(containing the incoming and scattered electron momentum vectors) and
the reaction plane (containing the momentum transfer $(\vec{q}\,)$ and
outgoing proton momentum vectors.)  $\sigma_M$ is the Mott
cross-section, $f _{rec}=\left| {1+{{\left({\omega p_p-E_pq\cos \left(
{\theta _{pq}} \right)} \right)}
\mathord{\left/ {\vphantom {{\left( {\omega p_p-E_pq\cos \left(
{\theta _{pq}} \right)} \right)} {M_Ap_p}}}
\right. \kern-\nulldelimiterspace} {M_Ap_p}}} \right|$ 
is the recoil function and
$C_{kin}={{M_{rec}M_pp_p} \mathord{\left/ {\vphantom {{M_{rec}M_pp_p}
{M_A}}} \right. \kern-\nulldelimiterspace} {M_A}}$ 
is a kinematic factor\cite{10}, where $M_{rec}$ is 
the recoil mass, $E_p$, $p_p$
and $M_p$ are the proton energy, momentum and mass respectively. The
momentum and energy transfer are given by $q$ and $\omega$
respectively, and $\theta_{pq}$ is the angle between the momentum
transfer and the outgoing proton. For $\phi_{pq}=0$ and $\pi$, the
proton is said to be emitted "in plane". To separate the various
structure functions, several measurements are needed at the same $q$ and
$\omega$. The $R_{L}$ and $R_{T}$ structure functions can be separated by a
Rosenbluth separation. The $R_{TT}$ structure function can be separated by
an out of plane measurement, and the $R_{LT}$ structure function can be
separated by measuring the cross-section at both sides of the $q$ vector
($\phi_{pq} =0$ and $\phi_{pq} = \pi$) leaving all other kinematic 
variables the same\cite{11}. $R_{LT}$ is then determined from:
\begin{eqnarray}
R_{LT}&=&{1 \over {2v_{LT}}}{{\left( {2\pi } \right)^3} \over
      {C_{kin}\sigma _{M} f_{rec}^{-1}}} \times \cr\hfill
    & & \left\{ 
{{d^5\sigma \left( {\phi_{pq}=0}    \right)} 
	\over {d\omega d\Omega _ed\Omega _p}}-
 {{d^5\sigma \left( {\phi _{pq}=\pi }\right)} 
	\over {d\omega d\Omega _ed\Omega _p}} 
	 \right\} 
\label{eq:rlt}
\end{eqnarray}
The kinematics of in-plane coincidence scattering allows for only four
independent scalar variables. This set of variables can be chosen to
be $\left\{ {q,\omega ,E_p,\theta _{pq}} \right\}$ or $\left\{
{q,\omega ,E_m,P_m} \right\}$ (among other possibilities.)  The
variable $\theta_{pq}$, therefore, is not independent from the missing
energy $\left( {E_m\equiv \omega -T_p-T_R}
\right)$ and the missing momentum ($P_m\equiv \left|
{\vec{p_p}-\vec{q}\,}
\right| =\sqrt {p_p^2+q^2-2p_pq\cos \left( {\theta _{pq}} \right)}$.)\hfill
Thus the range in missing momentum that is sampled in the
measurement depends directly on the range of the opening angle,
$\theta_{pq}$.

\section{The Experiment and Analysis}

The experiment was performed in the North Hall of the MIT-Bates linear
electron accelerator facility.  The electron beam energy was $576.0
\pm 0.9\,MeV$ at an average current of about $3\,\mu{A}$ 
and a $1\%$ duty factor.  The
target was a $208.9\,mg/cm^2$ thick carbon film. The electrons were
detected at $44^\circ$ using the high resolution (${{\Delta p}
\mathord{\left/ {\vphantom {{\Delta p} {p\approx }}}
\right. \kern-\nulldelimiterspace} {p\approx }}10^{-4}$) energy loss
spectrometer system\cite{12} (ELSSY) in coincidence with in-plane protons
detected by a prototype "out-of-plane spectrometer" 
(OOPS)\cite{13,14,15,16,17,18}. 
These settings correspond to quasielastic kinematics with a central momentum
transfer of $q=404\,MeV/c$ and energy transfer 
of $\omega=112\,MeV$. The protons
were detected at two successive settings of $42.9^\circ$ 
and $64.7^\circ$ (on the
opposing side of the beam line from the electron spectrometer)
corresponding to opening angles of 
$\theta_{pq}=10.2^\circ$ $(\phi_{pq}=0^\circ)$ 
and $\theta_{pq}=11.7^\circ$ $(\phi_{pq}=180^\circ)$  respectively. 
The maximum opening angle was limited by geometric constraints of the 
North Hall. A slight mismatch of $\theta_{pq}$ between the two 
settings was caused by an
initial ambiguity in the beam energy. This was fully resolved after
the experiment and accounted for in the data analysis.  Data were
collected for proton momenta from $326\,MeV/c$ to $462\,MeV/c$ by stepping
through four sets of central momenta for the OOPS at each of the two
angular settings. This yielded data with a missing energy range
covering both p- and s-shell proton knockout ($E_{m}=0$ to $65\,MeV$)
from the carbon nucleus.

In order to make a meaningful extraction of $R_{LT}$ using 
equation \ref{eq:rlt}, all
four independent kinematic variables for $\phi_{pq}=0^\circ$ and  
$\phi_{pq}=180^\circ$ data must be identical. This subtraction 
of the two measurements is complicated by 
the finite acceptance of the spectrometers. To maximize the accuracy
of the extraction the data were completely analyzed in two dimensions,
using bins of missing momentum vs. missing energy.  The experimental
cross-section for each bin
 $B\left( {E_m,P_m} \right)$ is then given by:
\begin{equation}
\left\langle {{{d^6\sigma } \over {dE_fdE_md\Omega _ed\Omega _p}}}
\right\rangle _B={{\sum\limits_r {n_rN_r\left( {E_m,P_m} \right)}}
\over {\sum\limits_r {v_rV_r\left( {E_m,P_m} \right)}}}
\label{eq:sigexp}
\end{equation}
The summation is over all runs that have data contributing to bin $B\left(
{E_m,P_m} \right)$. $N_r\left({E_m,P_m}\right)$ is the number of
counts contributing to this bin for run $r$, and $n_r$ is the normalization
for that run (see below). $V_r$ is the phase-space volume defined by:
\begin{equation}
V_r\left( {E_m,P_m} \right)_B=\int\limits_B {\varepsilon
_{A_r}}dE_fdE_md\Omega _ed\Omega _p
\label{eq:spvol}
\end{equation}
where $\varepsilon _{A_r}$ is the six dimensional efficiency of the
spectrometers (from a Monte Carlo simulation) to detect particles
contributing to a bin $B\left( {E_m,P_m} \right)$ and the integration
is over the acceptance of the spectrometers.  A two dimensional
contour plot (figure 1) of this volume for each of the two data sets
reveals the complicated nature of the acceptance and the need for a
mask to enforce overlap of the two data sets. This mask is chosen so
that the phase-space volume for each of the two settings is large
enough to expect a reasonable number of counts in the
$\left({E_m,P_m}\right)$ bins inside the masked region.
The phase-space volumes are normalized for each run by a factor
given by:
\begin{equation}
v_r={{C_r} \over e}{{N_At_r} \over {A\cos \left( {\theta _t}
\right)}}
\label{eq:vr}
\end{equation}
Where $C_r$ is the amount of charge onto the target, $t_r$
is the target thickness, $A$ is the atomic weight of the target 
and $\theta_t$ is
the target angle. $N_A$ and $e$ are Avogadro's number and the electron
charge respectively.

The data were corrected by the measured efficiencies of the
spectrometers, which is incorporated into the factor $n_r$ of formula
\ref{eq:sigexp}. The integrated efficiency of the electron spectrometer was
determined by measuring the elastic $^{1}H(e,e')$ cross-section and
comparing it with the values predicted by the Mainz\cite{19} parametrization of
the proton form factors\cite{17,18}. The integrated spectrometer efficiency
was found to be $98.4\pm{0.2}\%$. The integrated efficiency of the proton
spectrometer was determined by $^{1}H(e,e'p)$ scattering, and was found to
be $96.3\pm{1.0}\%$ after all known detector dead times ($4\%$ to $6\%$) 
were taken into account\cite{17,Jordan,18}. The data were also corrected for radiative
processes in a two dimensional manner using the code {RADC}\cite{20}, which
unfolds the radiative tails in the missing momentum versus missing
energy histogram using a formalism based on that of 
Mo and Tsai \cite{21,22,23,24}. 
These corrections were typically around $22\%$ for internal
bremstrahlung (Schwinger correction\cite{25}) and small 
($2\%$ to $3\%$) for all other
processes. After this correction the $R_{LT}$ response was calculated for
each bin, and the resulting histograms were projected onto the missing
energy axis for interpretation. Projection onto the missing momentum
axis was not fruitful due to the sparsity of the data, so only one
integrated data point could be obtained in the $P_m$ dimension.

\section{Results and Discussion}
Figure \ref{fig2} shows the histograms of the cross-section 
results for $\phi_{pq}=0^\circ$ and  $\phi_{pq}=180^\circ$, 
along with the extracted $R_{LT}$ values. The
data clearly exhibit the p-shell proton knockout region from about 
$13\,MeV$ to $28\,MeV$ which is known to contain several excited states of the
$^{11}B$ residual nucleus\cite{26}. The resolution in this 
experiment was not high
enough to distinguish these states, although there is some indication
of a peak corresponding to an excited state at about $24\,MeV$ in missing
energy ($8\,MeV$ excitation)\cite{Kester}. The integrated cross-sections over the
entire p-shell region are $8.3\pm 0.2\pm 0.1\, nbarn/(MeV sr^2)$ for 
$\phi_{pq}=0^\circ$ ($\theta_p=42.9^\circ$) and
$13.1\pm 0.3\pm 0.3\, nbarn/(MeV sr^2)$ for 
$\phi_{pq}=180^\circ$ ($\theta_p=64.7^\circ$). The s-shell region ranges
from $28\,MeV$ up to $50\,MeV$ in missing energy. This upper limit was
established by a previous experiment separating $R_L$ and 
$R_T$\cite{4,27} where it was
found that $R_L$ vanishes for missing energies larger than $50\,MeV$. The
integrated cross-sections for the s-shell region are 
$3.5\pm 0.2\pm 0.1\, nbarn/(MeV sr^2)$ for 
$\phi_{pq}=0^\circ$($\theta_p=42.9^\circ$) 
and $5.9\pm 0.2\pm 0.1\, nbarn/(MeV sr^2)$ for $\phi_{pq}=180^\circ$ 
($\theta_p=64.7^\circ$). The integrated cross-sections for the region above the
s-shell, from $50\,MeV$ to $75\,MeV$, are $0.8\pm 0.3\, nbarn/(MeV sr^2)$ 
for $\phi_{pq}=0^\circ$
and $1.7\pm 0.4\, nbarn/(MeV sr^2)$ for $\phi_{pq}=180^\circ$.

\begin{figure}
\epsfxsize=3.0in
\centerline{\epsfbox{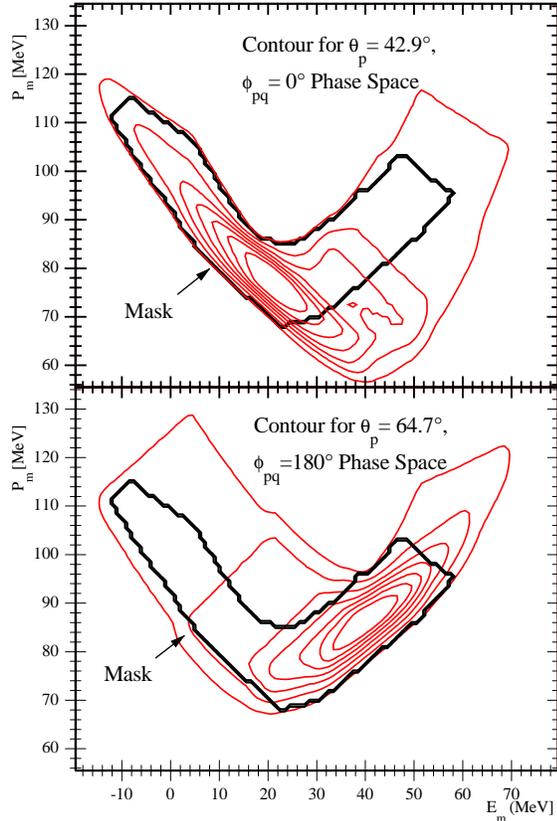}}
\caption{Contour histogram of the summed acceptance
volume $\sum{v_rV_r\left(E_m,P_m\right)}$ for the 
$\phi_{pq}=0^\circ$ and $\phi_{pq}=180^\circ$ data, and the overlap (bold line) of
the two regions that was used for the analysis of this data set (see text.)}
\label{fig1}
\end{figure}

\begin{figure}
\epsfxsize=3.0in
\centerline{\epsfbox{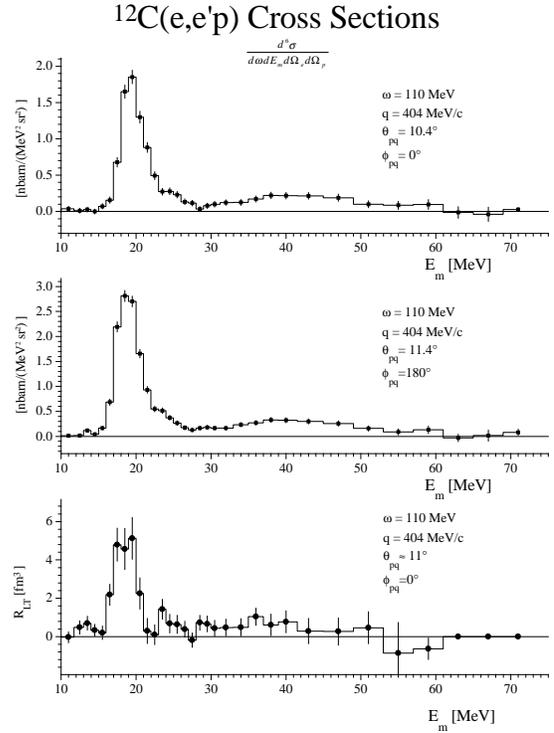}}
\caption{Cross-sections and extracted $R_{LT}$ response functions versus
missing energy. The data has been fully corrected for radiative
effects and normalizations. Error bars include the statistical
uncertainty only. }
\label{fig2}
\end{figure}

For extraction of $R_{LT}$, the data were
masked (see figure \ref{fig1}) before the $R_{LT}$ calculation and subsequent
projection onto the missing energy axis. The resulting histogram of
$R_{LT}$ (figure \ref{fig2}) clearly shows the p- shell and s-shell 
regions with
strengths of $25.\pm 2.\pm 2.fm^3$ and $14.\pm 2.\pm 1.fm^3$ 
respectively. The $R_{LT}$
strength is consistent with zero for missing energies larger than $50\,MeV$,
which is consistent with the findings of ref. \cite{4,27} that $R_L$ is
zero for missing energies larger than $50\,MeV$.  

\begin{figure*}
\epsfxsize=6.0in
\centerline{\epsfbox{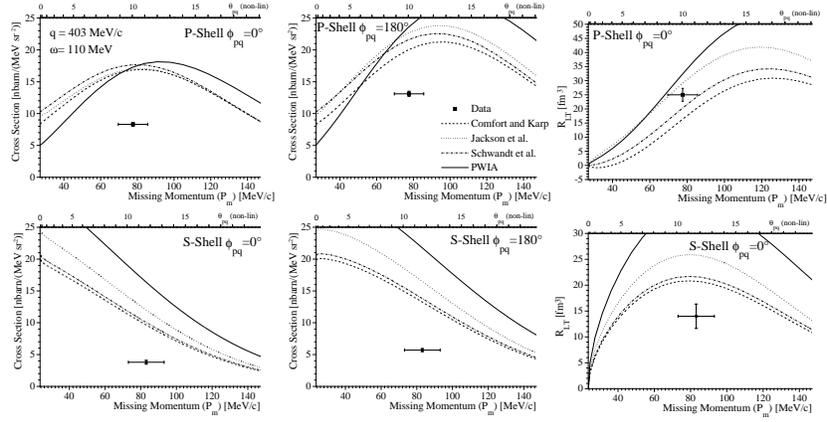}}
\caption{Data compared with DWIA calculations
[28,29]. 
The top row shows the results for the p-shell \\
$(13\, MeV <$ $E_m$ $< 28\, MeV)$ for $\phi_{pq}=0^\circ$,
$\phi_{pq}=180^\circ$ and $R_{LT}$, the bottom row shows the same for the
s-shell $(28\, MeV <$ $E_m$ $< 50\, MeV)$. The various curves are for different
optical potentials used for the final state proton distortions (see
text.)}
\label{fig3}
\end{figure*}

The experimental
results are compared with non- relativistic distorted wave impulse
approximation (DWIA) calculations\cite{28,29} in figure \ref{fig3}. 
These calculations
use a Wood Saxon potential well to calculate the bound state wave
functions, and an optical potential model to calculate the distortion
of the outgoing protons. The parametrizations of the optical
potentials used were due to Comfort {\em et al.}\cite{30}
(dashed lines), Jackson 
{\em et al.}\cite{31} (dotted lines) and Schwandt {\em et al.}\cite{32}
(dot-dash lines). Plane wave
impulse approximation (PWIA) calculations (solid lines) are also shown
for comparison. 
A disscussion of these optical 
potentials can be found in ref. \cite{Steenhoven}.

There are significant discrepancies between the DWIA
calculations and the data for both the p- and s-shell
cross-sections. These discrepancies can be expressed as spectroscopic
factors, which are defined as the ratio of the experimental
cross-sections to the theoretical cross-sections assuming one proton
per shell.
The nominal value for the p-shell (s-shell) spectroscopic factor would
thus be 4 (2), however it has been well documented that these factors 
are closer to 2.5 and 0.9 for experimental data.\cite{Lapikas}
Taking the
averaged result of the cross-sections calculated using the three
different optical potentials, the spectroscopic factors for the
p-shell are $1.93\pm0.07$ at $\phi_{pq}=0^\circ$ and 
$2.40\pm0.14$ at $\phi_{pq}=180^\circ$, and for
the s- shell $0.67\pm0.09$ at $\phi_{pq}=0^\circ$ and 
$0.82\pm0.08$ at $\phi_{pq}=180^\circ$. The
uncertainties quoted include the full experimental uncertainties and
estimated theoretical uncertainties equal to the standard deviations
of the theoretical predictions added in quadrature. These
spectroscopic values fall well below the expected values of 4 and 2,
but are consistent with previous experiments which have found spectral
functions in the range of 2 to 3 for the p-shell and 0.6 to 1.2 for
the s-shell\cite{33}.  

An extension of the spectroscopic factor can be
constructed by comparing the results of the extracted $R_{LT}$ value with
the theoretical $R_{LT}$ predictions. This comparison results in
$R_{LT}$-spectroscopic factors of $4.6\pm1.1$ for the p-shell and 
$1.4\pm0.3$ for the s-shell. 
This is in surprisingly good agreement with the nominal
values of 4 and 2. Since $R_{LT}$ is the difference of the 
$\phi_{pq}=0^\circ$ and
$\phi_{pq}=180^\circ$ cross-sections, it should be expected 
to scale with the same
spectroscopic factors as the full cross-sections. This discrepancy
indicates that DWIA calculations are not sufficiently accurate to
predict the separated $R_{LT}$ structure function. A similar conclusion was
reached by Spaltro {\em et al.}\cite{Spaltro} who found that the $R_{LT}$ 
spectroscopic factor
for p-shell proton knockout from $^{16}O$ in quasielastic kinematics was
enhanced as compared to the $R_{L}$ and $R_T$ spectroscopic factors.  

An additional, more complete calculation was performed for the s-shell
region by J. Ryckebush in the framework of a self consistent
Hartree-Fock Random Phase Approximation (HF-RPA)\cite{35,36}. This calculation
included two body currents related to meson exchange currents and
$\Delta(1232)$ excitation. They also included two nucleon knockout
contributions to the cross-section. For the distribution of the
s-shell strength in missing energy, a Lorentzian curve was used:
\begin{eqnarray}
\lefteqn{S_{s{1 \over 2}}\left( {E_m} \right)= }  \nonumber  \\
& & {1 \over \pi}{{W\left( {E_m-\varepsilon _F} \right)} \over {\left(
{E_m-\varepsilon _F-\left| {\varepsilon _h-\varepsilon _F} \right|}
\right)^2+\left[ {W\left( {E_m-\varepsilon _F} \right)} \right]^2}}
\label{eq:s}
\end{eqnarray}
which includes an energy-dependent width as prescribed by Mahaux\cite{37}:
\begin{equation}
W\left( E \right)={{9E^4} \over {E^4+\left( {13.27}
\right)^4}}\left[ {MeV} \right]
\label{eq:w}
\end{equation}
which was obtained from a
compilation of experimental data\cite{38}. Furthermore $\varepsilon_h$
is the energy needed
to remove a proton from the s-shell and $\varepsilon_F$ is the Fermi energy (
$\varepsilon_F\approx18\,MeV$,
$\left|{\varepsilon _h-\varepsilon _F} \right|\approx 23\,MeV$, 
for a $^{12}C 1s_{1/2}$ proton).

\begin{figure}
\epsfxsize=3.0in
\centerline{\epsfbox{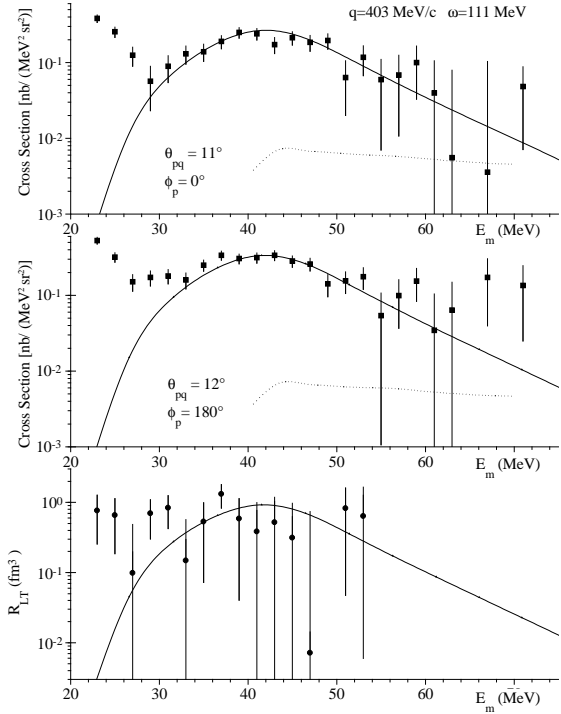}}
\caption{Comparison of the s-shell data with an HF-RPA
calculation [36] using a spectroscopic function to model the shape of
the s-shell. The dotted line in the top two plots indicate the
contribution from two-nucleon knockout.  }
\label{fig4}
\end{figure}

The cross-section versus missing energy is then given by 
\begin{equation}
\sigma _{1S{\textstyle{1 \over 2}}}\left( {E_m} \right)=0.3\cdot
S_{S{\textstyle{1 \over 2}}}\left( {E_m} \right)\sigma
_{1S{\textstyle{1 \over 2}}}^{HF-RPA} 
\label{eq:sigs}
\end{equation}
which includes a reduction factor of 0.3. This factor is similar
to a spectroscopic factor, but is smaller on account of the 
lack of an absorptive immaginary potential in the
treatment of the final state interactions in the HF-RPA
calculation \cite{R_PRIV}\cite{Sluys}.
 The same factor was used for both the cross-sections and
the extracted $R_{LT}$ structure function. The resulting curves in
figure \ref{fig4} show a good agreement between theory and data in the
s-shell region. This indicates that more complete calculations are
needed to predict the $R_{LT}$ structure functions and that the Mahaux
shape models the s-shell distribution well. This better agreement
might be due to the sensitivity of the $R_{LT}$ structure function to
two body currents in the nucleus, as suggested by ref. \cite{39}.
Figure \ref{fig4} also shows the contribution of two nucleon knockout
to the cross-section.  This contribution is too small to explain the
difference between the two calculations. Also note that since this
data was taken at an electron scattering angle of $44^\circ$ the
virtual photon was mostly longitudinally polarized and therefore
little strength is expected in the continuum ($E_m> 50\,MeV$), based
on the findings of refs. \cite{4,27}.

\begin{figure}
\epsfxsize=3.0in
\centerline{\epsfbox{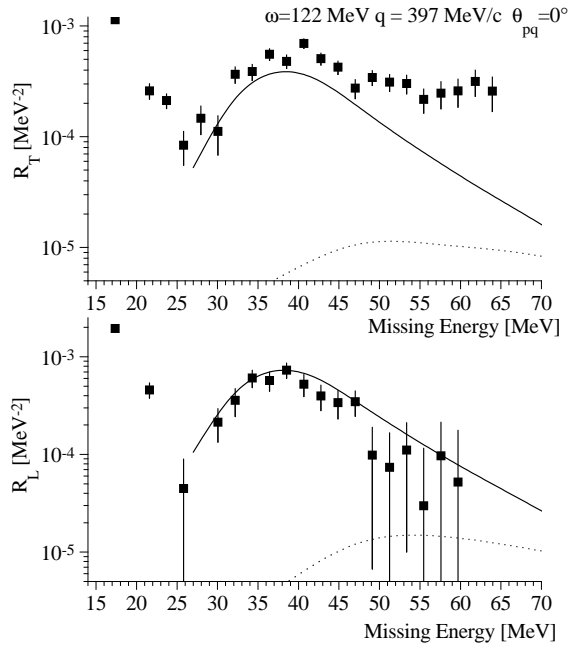}}
\nopagebreak
\caption{Comparison of the
separated $R_L$ and $R_T$ structure functions from Ulmer {\em et al.} [4], with an
HF-RPA calculation [36] using an s-shell spectroscopic function. The
dashed line indicates the two-nucleon knockout contribution.}
\label{fig5}
\end{figure}

A different picture is obtained when the same calculation is compared to
$R_L$ and $R_T$ structure functions separated by Ulmer {\em et al.} \cite{4}. 
These quasi-elastic data were taken in parallel kinematics ($\theta_{pq}=0$)
at an
energy transfer $\omega=122\,MeV$ and a momentum transfer $q=397\,MeV/c$. The
calculation was scaled by a factor of 0.4 to fit the $R_L$ data (instead
of the 0.3 used above). As can be seen in figure \ref{fig5}, the shape of $R_L$
for the s- shell data is correctly described by the calculation, but
for $R_T$ the calculation underestimates the data. Also the continuum
strength for $R_T$ is significantly under predicted by this
calculation. This discrepancy cannot be explained by the two body
contributions to the cross-section or the two-nucleon knockout
contribution (dashed line). We must conclude that even the HF-RPA
calculations cannot describe the separated transverse response
function consistently. The good description of the $R_{LT}$ data may be due
to the combination of the longitudinal response going to zero (and
therefore driving $R_{LT}$ to zero also) and the chosen longitudinal
electron kinematics.  

\section{Summary}
A first measurement of the $R_{LT}$ structure function for the carbon
nucleus at quasi-elastic kinematics was performed. This measurement
verifies that the s-shell strength vanishes for missing energies
larger than $50\,MeV$. The strength of the $R_{LT}$ structure function
cannot be correctly described by a DWIA calculation when using the
same spectroscopic factors that are needed to predict the
cross-sections. This discrepancy may be due to the sensitivity of
$R_{LT}$ to the two body contributions to the nuclear current, as
indicated by a more complete HF-RPA calculation, which successfully
predicts $R_{LT}$ and $R_L$. However, these same calculations do not
correctly replicate the measured $R_T$ structure
function. Spectroscopic functions by Mahaux accurately describe the
s-shell distributions in missing energy but they underestimate the
strength in the continuum $(E_m>50\,MeV)$.  Additional measurements
and better calculations are needed to fully understand the
sensitivities of the $R_{LT}$ and other structure functions to the
various components of the nuclear current. A consistent set of
measurements that separate as many response functions as possible will
go a long way towards the understanding of the nuclear current. A
cluster of OOPS spectrometers is currently being used at the Bates
accelerator laboratory to make such measurements\cite{40}.

\section{Acknowledgments}
  We would like to thank the
technical staff of the Bates Laboratory for their assistance in
carrying out this work. We also would like to thank T. W. Donnelly and
J. Ryckebusch for their comments and assistance.  This work was
supported in part by the National Science Foundation under grants
no. NSF PHY 89-21146 and no. NSF PHY 92-00435 and the US Department of
Energy under contracts no. DE-AC02- 76ERO3069 and
no. DE-FG-2-88ER40415.

%
%

\onecolumn
\twocolumn

\begin{references}

%
\bibitem[*]{UNH}	  Present address:
			  University of New Hampshire,
			  Durham, NH03824.
\bibitem[*\ddag]{TRIUMF}  Present address:
			  TRIUMF, Vancouver, BC, Canada.
\bibitem[\dag]{CEBAF}     Present address:
                          TJNAF,
                          Newport News, VA 23606.
\bibitem[\ddag]{UKL}      Present address:
                          University of Kentucky,
                          Lexington, KY  40506.
\bibitem[\S]{AZST}        Present address:
                          Arizona State University,
                          Tempe, AZ 85287.
\bibitem[\S\S]{MGH}       Present address:
                          Massachusetts General Hospital
                          NMR Center,
                          Charlestown, MA 02129.
\bibitem[\dag\dag]{UATH}  Present address:
                          National and Capodistrian
                          University of Athens,
                          Athens, Greece.
\bibitem[\ddag\ddag]{ODU} Present address:
                          Old Dominion University,
                          Norfolk, VA 23529.
\bibitem[\S\dag]{FIU}     Present address:
		          Florida International University,
                          Miami, Florida FL 33199
\bibitem[\S\ddag]{UVA}    Present address:
		          University of Virginia,
                          Charlottesville, Virginia 22903
\bibitem[\S*]{KNU}        Present address:
                          Kyungpook National Univ, 
			  Taegu 702-701, South Korea
\bibitem[\dag*]{UW}       Present address:
		          University of Washington,
			  Seattle, WA 98195   
\bibitem[**]{SUNY}	  Present address:
			  SUNY Stony Brook,
			  Stony Brook, NY 11790	


\bibitem[1]{1} For an overview see "Modern Topics in Electron Scattering", 
edited by B. Frois and I. Sick, World Scientific, ISBN 9971-50-975. (1991).

\bibitem[2]{2} Jean-Marc Laget, "Electrodisintegration of 
Few-Nucleon Systems." ,
 and W. Bertozzi, R. W. Lourie, and E. J. Moniz, 
"Nuclear Response Functions at Large Energy and Momentum Transfer",
in Ref\cite{1}.

\bibitem[3]{Lapikas} L. Lapikas, Nucl. Phys. {\bf A 553}, 297c (1993).
\bibitem[4]{3} R. R. Whitney {\em et al.}, Phys. Rev. {\bf C 9}, 2230 (1974).
\bibitem[5]{4} P. Ulmer {\em et al.}, Phys. Rev. Lett., {\bf 59}, 2259 (1987).
\bibitem[6]{5} L. B. Weinstein {\em et al.}, Phys. Rev. Lett. {\bf 64}, 1646 (1990).
\bibitem[7]{6} R. W. Lourie {\em et al.}, Phys. Rev. Lett. {\bf 56}, 2364 (1986).
\bibitem[8]{7} H. Baghaei {\em et al.}, Phys. Rev. {\bf C 39}, 177 (1989).
\bibitem[9]{Kester} L.J.H.M. Kester, H.P. Blok, W.H.A. Hesselink, A. Pellegrino, E. Jans, L. Lapikas, G. van der Steenhoven, A. Zondervan, J. Ryckebusch, Phys. Lett. {\bf B 366},44 (1996).
\bibitem[10]{Kester2} L.J.H.M. Kester {\em et al.}, Phys. Lett. {\bf B 344}, 79 (1995).
\bibitem[11]{8} J. M. Le Goff {\em et al.}, Phys. Rev. {\bf C 55}, 1600 (1997).
\bibitem[12]{Spaltro} C. M. Spaltro {\em et al.}, Phys. Rev. {\bf C 48}, 2385 (1993).
\bibitem[13]{Jordan} D. Jordan {\em et al.}, Phys. Rev. Lett. {\bf 76}, 1579 (1996).
\bibitem[14]{Ducret} J. E. Ducret {\em et al.}, Phys. Rev. {\bf C 49}
\bibitem[15]{Kasdorp} W.J. Kasdorp {\em et al.}, Phys. Lett. {\bf B 393}, 42 (1997).
\bibitem[16]{Frommberger} F. Frommberger {\em et al.}, Phys. Lett. {\bf B 339}, 17 (1994).
\bibitem[17]{9} T. W. Donnelly and A. S. Raskin, Ann. Phys., {\bf 169},247 (1986).

\bibitem[18]{10} A slightly different formalism due to Boffi et. al. 
Nucl. Phys. {\bf A 435}, 697 uses the constant: The difference in the 
formalisms is absorbed in the details of the structure 
functions.

\bibitem[19]{11} C. N. Papanicolas {\em et al.}, Nucl. Phys. {\bf A 497}, 509c (1989).
\bibitem[20]{12} W. Bertozzi {\em et al.}, Nucl. Inst. Methods Phys. Res. {\bf 162}, 211 (1979).
\bibitem[21]{13} S. Dolfini {\em et al.}, Nucl. Inst. Methods Phys. Res. Sec. {\bf A 344}, 571 (1994).
\bibitem[22]{14} J. Mandeville {\em et al.}, Nucl. Inst. Methods Phys. Res. Sec. {\bf A 344}, 583 (1994).
\bibitem[23]{15} J. Mandeville {\em et al.}, Phys. Rev. Lett. {\bf 72}, 3325 (1994).
\bibitem[24]{16} S. Dolfini {\em et al.}, Phys. Rev. {\bf C 51}, 3479 (1995).
\bibitem[25]{17} D. Jordan, Ph.D. Thesis MIT-Bates (1994) unpublished,
\bibitem[26]{18} M. Holtrop, Ph.D. thesis MIT-Bates (1995) unpublished.
\bibitem[27]{19} G. Simon {\em et al.}, Nucl. Phys. {\bf A 333}, 381 (1980).
\bibitem[28]{20} E. Quint, Ph.D. thesis NIKHEF (1988) unpublished. 
\bibitem[29]{21} L. W. Mo and Y. S. Tsai, Rev. Mod. Phys. {\bf 122}, 1898 (1961) and Rev. Mod. Phys. {\bf 41}, 208 (1969).
\bibitem[30]{22} S. Penner, Nuclear Structure Physics, Proceedings of the 18th Scottish Univ. Summer School in Physics, p 284 (1977).
\bibitem[31]{23} J. Friedrich, Nucl. Inst. Methods, {\bf 129}, 505 (1975).
\bibitem[32]{24} D.J.S. Findlay and A. R. Dusautoy, Nucl. Inst. Methods, {\bf 174}, 531 (1980).
\bibitem[33]{25} J. Schwinger, Phys. Rev. {\bf 76}, 760 (1949).
\bibitem[34]{26} G. van der Steenhoven {\em et al.} Nucl. Phys. {\bf A 484}, 445 (1988).
\bibitem[35]{27} P. Ulmer, Ph.D. Thesis. MIT-Bates (1987) unpublished.
\bibitem[36]{28} C. Giusti and F. D. Pacati, Nucl. Phys., {\bf A 485}, 461 (1988).
\bibitem[37]{29} Distorted Wave (e,e'p) (DWEEPy) code, G. van der Steenhoven, NIKHEF.
\bibitem[38]{30} J. R. Comfort and B. C. Karp, Phys. Rev. {\bf C 21}, 2162 (1980).
\bibitem[39]{31} D. F. Jackson and I. Abdul-Jalil, J. Phys., {\bf G 6}, 481 (1980).
\bibitem[40]{32} P. Schwandt {\em et al.}, Phys. Rev. {\bf C 26}, 55 (1981).
\bibitem[41]{Steenhoven} G. van der Steenhoven {\em et. al.}, Nucl. Phys. {\bf A 480}, 547 (1988).
\bibitem[42]{33} W. Bertozzi, Nucl. Phys. {\bf A 527} ,347c (1991).
\bibitem[43]{35} J. Ryckebush, Ph.D. Thesis and private communications.
\bibitem[44]{R_PRIV} J. Ryckebusch private communication.
\bibitem[45]{Sluys} V. Van der Sluys, J. Ryckebusch, and M. Waroquier, Phys. Rev. {\bf C 54}, 1322 (1996).
\bibitem[46]{36} J. Ryckebusch {\em et al.} Phys. Rev. {\bf C 49 N5}, 2704 (1994).
\bibitem[47]{37} C. Mahaux {\em et al.}, Phys. Rep. {\bf 120}, 1 (1985).
\bibitem[48]{38} J. P. Jeukenne and C. Mahaux, Nucl. Phys. {\bf A 394}, 455 (1983).
\bibitem[49]{39} V. Van der Sluys, J. Ryckebusch and M. Waroquier, Phys. Rev. {\bf C 49}, 2695 (1994).
\bibitem[50]{40} Bates Experiment 94-08, ``A Measurement of the Structure Functions via Quasielastic$^{12}C$, $^{16}O(\vec{e},e'p)$'', W.-Y.\ Kim and C.\ Papanicolas (co-spokespersons).



\end{references}
\end{document}